\documentclass[10pt,twocolumn,twoside]{IEEEtran}
\usepackage{amsmath,amsfonts}
\usepackage{amssymb}
\usepackage{algorithmic}
\usepackage{algorithm}
\usepackage{array}
\ifCLASSOPTIONcompsoc
 \usepackage[caption=false,font=normalsize,labelfont=sf,textfont=sf]{subfig}
\else
 \usepackage[caption=false,font=footnotesize]{subfig}
\fi
\usepackage{textcomp}
\usepackage{stfloats}
\usepackage{url}
\usepackage{verbatim}
\usepackage{graphicx}
\usepackage{cite}

 \usepackage{booktabs}
\usepackage{amsthm}
\usepackage{multirow}
\usepackage{subfloat} 
 \usepackage{epstopdf}
 \usepackage{makecell}

\newtheorem{remark}{Remark}
\newtheorem{definition}{Definition}
\newtheorem{theorem}{Theorem}
\newtheorem{lemma}{Lemma}

\newtheorem{problem}{Problem}
\newtheorem{assumption}{Assumption}

\begin{document}

\title{On Game based Distributed Approach for General Multi-agent Optimal Coverage with Application to UAV Networks}

\author{Zixin Feng,~\IEEEmembership{Member,~IEEE,}
         Wenchao~Xue,~\IEEEmembership{Member,~IEEE,}
        Yifen Mu,~\IEEEmembership{Member,~IEEE,}
        Ming Wei,~\IEEEmembership{Member,~IEEE, }
        Bin Meng,~\IEEEmembership{Member,~IEEE}
        ~and Wei Cui,~\IEEEmembership{Member,~IEEE}

\thanks{\noindent Manuscript received xxx; revised xxx. This work was supported by the National Key R\&D Program of China (No. 2022YFA1004703), the National Natural Science Foundation of China (No. 62122083), and Chinese Academy of Sciences Youth Innovation Promotion Association (No. 2021003). }
\thanks{Zixin Feng, Wenchaoxue and Yifen Mu are with School of Mathematical Sciences, University of Chinese Academy of Sciences, Beijing, China and Academy of Mathematics and Systems Sciences, Chinese Academy of Sciences, Beijing, China. \itshape(Corresponding author: Wenchao Xue. Email: wenchaoxue@amss.ac.cn).}
\thanks{Ming Wei and Wei Cui are with South China University of Technology,  Guangdong, China. Bin Meng is with China Academy of Space Technology, Beijing, China.  }

}

\markboth{Manuscript for IEEE Transactions on Control of Network Systems, November~2025}%
 {Shell \MakeLowercase{\textit{et al.}}: A Sample Article Using IEEEtran.cls for IEEE Journals}


\maketitle
\begin{abstract}
This paper focuses on the optimal coverage problem (OCP) for multi-agent systems with a decentralized optimization mechanism. A game based distributed decision-making method for the multi-agent OCP is proposed to address the high computational costs arising from the large scale of the multi-agent system and to ensure that the game's equilibrium achieves the global performance objective's maximum value. In particular, a distributed algorithm that needs only local information is developed and proved to converge to near-optimal global coverage.
Finally, the proposed method is applied to maximize the coverage area of the UAV network for a target region. The simulation results show that our method can require much less computational time than other typical distributed algorithms in related work, while achieving a faster convergence rate. Comparison with centralized optimization also demonstrates that the proposed method has approximate optimization results and high computation efficiency.
\end{abstract} 

\section{Introduction}

The optimal coverage problem (OCP) for multi-agent systems has become a hot topic in many fields, such as robotics\cite{karapetyan2017efficient},
unmanned systems\cite{huang2022deployment} and satellite constellations\cite {chen2021general,chen2023vleo}. This problem is a classic optimization problem that aims to achieve maximum coverage or minimum coverage cost of the target under certain constraint conditions to improve the system's overall efficiency. 
Traditional methods for solving the OCP rely on centralized method\cite{ulybyshev2008satellite,mannadiar2010optimal,zhang2018fast,sun2023optimal,10876799}.  However, such methods often face problems such as high computational complexity and poor real-time performance when dealing with large-scale, dynamically changing multi-agent systems.
With the rapid development of multi-agent systems, their scale has continuously increased, and the distributed decision method has become an important tool for solving the OCP of large-scale multi-agent systems because of its scalability, robustness, and efficiency.
Zou and Zhang \cite{10829684} presented a hierarchical coverage control framework for multi-agent systems in porous environments, using a Voronoi-based sweep-and-assign scheme and load-balancing subdivision for optimal coverage deployment.
In \cite{abdulghafoor2021two}, a two-level distributed control framework for a multi-agent network is introduced to address dynamic coverage problems.  Dong and Li\cite{dong2023energy} presented a heuristic energy-efficient sensor deployment strategy for optimal coverage of underwater events. The strategy considered the surrounding event information, neighbor information, and congestion level to guide the sensor nodes to optimal positions. 
In \cite{9904913}, a distributed multi-robot coverage control algorithm for nonconvex environments was proposed, achieving the first constant-factor approximation guarantee for optimal coverage, with convergence to near-optimal solutions in polynomial time.

Learning-based methods are also applied in the research of the optimal coverage problem because they can continuously improve performance through iterative learning processes. 
Xiao and Wang\cite{xiao2020distributed} studied a distributed multi-agent dynamic area coverage algorithm based on reinforcement learning under different communication conditions. 
Meng and Kan\cite{meng2021deep} proposed a model-free framework based on deep reinforcement learning for dynamic coverage control of multi-agent systems, which allows agents to learn optimal control strategies directly from their interactions with the environment.
Wang and Fu\cite{wang2024distributed} studied the coverage control problem of multi-agent systems in a task region with unknown observation density functions and an integrated coverage performance function. An online distributed estimation algorithm is employed to learn the unknown density function and a distributed control law is developed to find the locally optimal coverage configuration.
However, the learning-based method faces difficulty in obtaining training data, which limits its application in practice.

Game based method is a powerful tool for distributed decisions. Each player has its utility function and tries to maximize it while considering the strategies of other players, which naturally fits the distributed decisions of multi-agent systems.  Thus, there has been a lot of literature using the games based method to study distributed OCP\cite{yazicioglu2016communication,trotta2018joint, hu2023multi,sun2017time,sun2021game,nemer2020game,gao2022coverage,varposhti2020distributed}. For example,
Yasin and Egerstedt\cite{yazicioglu2016communication} studied a communication-free algorithm for the distributed coverage of an arbitrary network by a group of agents with local sensing capabilities. Each agent aims to explore the maximizing coverage of a given graph and optimize their locations only by their own sensory inputs.
Sun and Wang\cite{sun2021game} proposed a game-theoretic approach to task allocation in multi-satellite systems and a distributed task allocation algorithm is given, in which restricted greedy, finite memory learning rules and the concept of "innovator" are used to enhance the efficiency of the algorithm. 
Nemer and Sheltami\cite{nemer2020game} studied a game-theoretical approach for the efficient deployment of agents in a multi-level and multi-dimensional assisted network. Each agent tries to achieve the best coverage based on its neighbors and the selected action.
Gao and Liang\cite{gao2022coverage} proposed a cooperative wireless coverage algorithm based on potential games for a multi-agent system, considering the stochastic wireless link failures caused by channel fading and noise in agent-to-agent communication links.

The optimal coverage problem also has many applications in the unmanned aerial vehicle (UAV) network. 
For example, 
In \cite{zhang2024distributed}, a method based on network evolutionary potential game theory was proposed for the dynamic task allocation problem of UAVs in complex and uncertain scenarios. It is proven that the optimal solution of the task allocation problem was a pure strategy Nash equilibrium of a finite strategy game.
Ma and Li\cite{ma2023dynamic} studied the dynamic tracking coverage problem of UAVs on unmanned ground vehicles and designed a tracking coverage algorithm based on $k$-means, which can adjust the number of UAVs reasonably according to the distribution of unmanned ground vehicles to ensure coverage effect.
Alzenad and El-Keyi\cite{alzenad20173} considered how to achieve three-dimensional optimized deployment of drones as base stations while meeting service quality requirements, to maximize user coverage and minimize transmission power.
In \cite{zhao2021multi}, Zhao and Liu studied the issue of multiple drones performing trajectory planning in unknown environments to achieve energy-efficient content coverage. A distributed reinforcement learning algorithm was proposed and effectively improved energy efficiency and content coverage.

It can be found that the existing works (for example,\cite{varposhti2020distributed,nemer2020game,gao2022coverage,sun2021game,zhang2024distributed}) lack some considerations. Firstly,
the problem models are only tailored to specific scenarios. Secondly, the majority of studies only consider a discrete strategy space. These features limit the generalizability. Finally, there is a lack of simultaneous consideration of observation performance and energy consumption in many articles, which leads to the limitation of practicability. As a comparison, this paper studies a game model that can be applied to a class of typical scenarios and considers both observation performance and energy consumption with a continuous strategy space, and the results can be easily generalized to a discrete strategy space. The main contributions of this paper are as follows.

\begin{enumerate}
    \item A game based distributed approach is developed for a class of typical OCP with both coverage performance and energy consumption performance. It is shown that the extreme value of the global performance objective can be obtained by finding the equilibrium of the multi-agent game, which is established by proving the game is a potential game.  
    \item A distributed algorithm that only utilizes local information is designed to search for the equilibrium solution of the OCP efficiently. Particularly, novel mechanisms of $\epsilon$-innovator and of local equilibrium judgment are established to speed up the convergence and reduce the computational costs. It is proven that the convergence of the proposed algorithm can be guaranteed under some weak assumptions. 
    \item The proposed method is applied to an unmanned aerial vehicle (UAV) network optimal coverage problem, in which UAVs try to maximize the total coverage area for a target region while saving energy. The simulation results show that the proposed method can improve the solving efficiency of the multi-agent OCP compared to the typical distributed methods and achieve performance close to that of centralized methods.  
\end{enumerate}

The rest of this paper is organized as follows. Section~\ref{sec_relate} investigates the related works on game-based distributed optimal coverage problems (OCPs). Section~\ref{sec2} formulates the game-based distributed OCP.  Section~\ref{sec3} gives a distributed optimal coverage searching algorithm and the proof of convergence. Section~\ref{sec4} shows the application of the proposed method in the UAV network. Section~\ref{sec5} conducts simulations to demonstrate the effectiveness of the proposed method and Section~\ref{sec6} draws a conclusion. The key notations used in this article are listed in Table~\ref{tab1}. Besides, some proofs are given in \ref{appen3}, \ref{appen1} and \ref{appen2}.
\begin{table}[htbp]
\renewcommand{\arraystretch}{1.5}
\caption{Key notations\label{tab1}}
\centering
\begin{tabular}{cc}
\hline
 Symbols & Notations\\
\hline
$a_k$ & The k-th agent\\
$\theta_k$ & The strategy of $a_k$\\
$\Theta_k$ & The strategy space $a_k$\\
$\theta$ & The strategy profile for all agents\\
$\Theta$ & The strategy profile space for all agents\\
$\theta_{-k}$ & The strategy profile for all agents except for $\theta_k$\\
$\mathcal{C}_k$ & The coverage of $a_k$ for the target\\
$\mathcal{N}_k$ & The index set of $a_k$'s neighbor\\
$N_k$ & The total number of $a_k$'s neighbors\\
$F$ & The global performance objective of the multi-agent system\\
$f_k$ & The local performance objective of $a_k$\\
$E_k$ & Energy penalty function of $a_k$\\
$\phi\left(\theta\right)$ & The potential function\\
$R_k$ & The regret values of $a_k$\\
$\tau_k$ & The observation function of $a_k$\\
 $\epsilon$ & Convergence accuracy of Algorithm \ref{DOCS}\\
$P$ & Total iteration number of Algorithm \ref{DOCS}\\
\hline
\end{tabular}
\end{table}

\section{Related work}\label{sec_relate}

Many studies had leveraged potential games to address multi-agent OCPs, particularly in domains such as mobile sensor networks \cite{sun2017time,varposhti2020distributed}, unmanned systems \cite{nemer2020game,gao2022coverage} and satellite resource allocation \cite{sun2021game}. However, the distributed OCPs for satellite constellation reconfiguration—specifically, maximizing observation time of the target—remain unexplored. 

In potential-game-based OCPs, spatial adaptive play (SAP), or log-linear learning (LL) in some articles, is a widely adopted method for achieving Nash equilibria. Specifically, SAP is a typical distributed random algorithm that selects a random agent to update its strategy following a Boltzmann distribution, while the remaining agents keep their current strategies. Further works of SAP are developed based on the above paradigm. For instance, in \cite{sun2017time} and \cite{zhang2024distributed}, a class of time-varying learning factors in the Boltzmann distribution was proposed and analyzed to balance exploration and optimization. In \cite{li2017potential} and \cite{wu2024joint}, a strategy update rule, called binary log-linear learning (BLL), was used to reduce computational burdens. Compared to LL, which updates the agent strategy from all strategy spaces at each iteration, BLL only allows the agent to update its strategy from the previous action and a trial action. However, SAP exhibits critical drawbacks. Firstly, its stochastic nature requires iterations to approach infinity for theoretical convergence guarantees, resulting in slow computation and poor real-time performance. Secondly, SAP relies on a central node to randomly select one agent for updates and then broadcast decisions globally, which is actually a pseudo-distributed manner. This centralized intervention is infeasible in many systems. Thirdly, these random algorithms often have parameters designed to trade off exploration and convergence. The design of these parameters relies on experience and is difficult to transfer between different problems. Finally, the computational efficiency of SAP is severely limited when dealing with large strategy sets.

Best-response-based method is another important approach for solving a potential game. Early works (\cite{ai2008optimality,yang2013towards}) used the best response rule (BRR), which randomly selected one agent to update its strategy to the best response, while other agents kept their current strategy. Because only one agent is allowed to update, the convergence speed of these methods is very slow. Distributed task allocation algorithm (DT2A) \cite{sun2021game} provides direct inspiration for the algorithm presented in this article. It proposed a distributed framework for satellite task allocation using an "innovator" mechanism to accelerate convergence and guarantee the distributed characteristics. In each iteration, agents identify innovators through local communication and only allow innovators to update strategies according to the best response. Additionally, it leveraged a memory mechanism in which each agent updates its strategy based on its memory series rather than on the best response. However, DT2A also suffered from limitations. Firstly, solutions are confined to finite action sets, failing to handle continuous optimization. Furthermore, \cite{sun2021game} did not analyze the upper bound on the number of iteration steps for convergence, which may lead to an overly conservative setting of the total number of iterations. Finally, all agents needed to compute their best responses in each iteration, incurring high overhead for large-scale systems.
\begin{table*}[!ht]
\footnotesize
\caption{Comparison of limitations in existing distributed methods for potential-game-based OCPs}
\label{tab:limit}
\centering
\tabcolsep 15pt 
\begin{tabular*}{\textwidth}{@{\extracolsep{\fill}}lp{0.6\textwidth}@{}}
\toprule
\textbf{Method} & \textbf{Main considerations} \\
\midrule
\begin{tabular}[c]{@{}l@{}}Spatial Adaptive Play (SAP)\\(e.g.,\cite{sun2017time,zhang2024distributed,gao2022coverage,wu2024joint}) \end{tabular} 
& \begin{itemize}
  \item Slow convergence due to stochastic updates, requiring infinite iterations for theoretical guarantees
  \item Pseudo-distributed architecture relying on a central node for agent selection
  \item Parameter sensitivity requiring manual tuning of exploration-convergence trade-off
  \item Limited computational efficiency when dealing with large strategy sets.
\end{itemize} \\
\begin{tabular}[c]{@{}l@{}}Best-Response-Based  Methods \\(e.g., BRR\cite{ai2008optimality,yang2013towards}, DT2A\cite{sun2021game})\end{tabular} 
& \begin{itemize}
  \item Low convergence efficiency from single-agent update mechanism (BRR)
  \item High computational overhead with all agents calculating best responses each iteration (DT2A)
  \item Lack of a theoretical upper bound for convergence iterations
  \item Inability to handle continuous optimization in strategy spaces (DT2A)
\end{itemize} \\
\bottomrule
\end{tabular*}
\end{table*}

The limitations in existing distributed methods for potential-game-based OCPs are shown in Table~\ref{tab:limit}.  To address the above gaps, this paper has the following advantages. First of all, a class of continuous OCPs with joint coverage-energy objectives is modeled and proved to be a potential game, unlike scenario-specific prior work. Second, we propose a distributed algorithm with the $\epsilon$-innovator mechanism and with a local equilibrium judgment mechanism for the potential game with continuous strategy space. The upper bound of the maximum iteration step for the algorithm's convergence is given theoretically. Finally, the simulation results further prove the effectiveness of the proposed method.

\section{Problem formulation}\label{sec2}
The interest of this paper is that each agent can make decisions in a distributed manner to achieve the extreme point of the global optimization objective. Because the goal is that agents make decisions independently based on their own utility functions and local information, it is natural to model distributed OCP using game based method. Therefore,  the multi-agent OCP is formulated in the frame of game theory in this section.

\subsection{Multi-agent optimal coverage problem}\label{sec2I}
\begin{figure}[!htbp]
\centering
\includegraphics[width=0.4\textwidth]{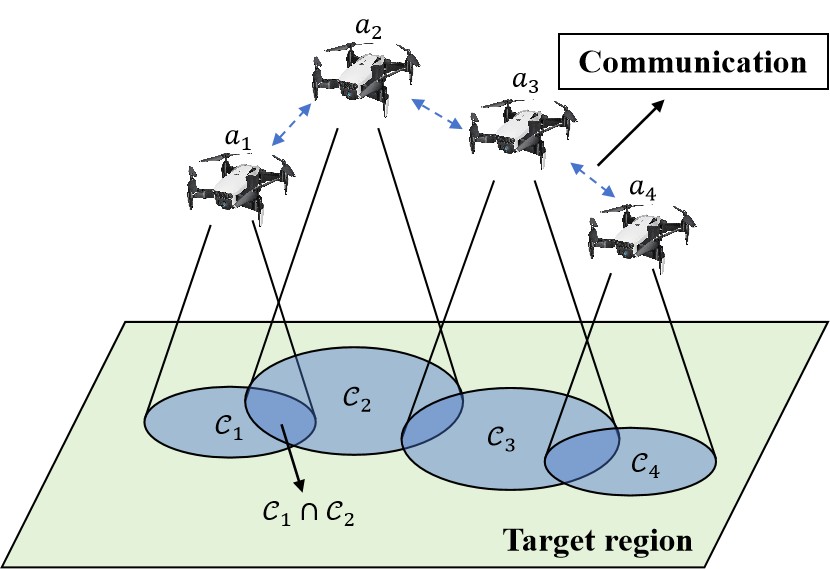}
\caption{Unmanned aerial vehicles coverage scenario. }
\label{fig2}
\end{figure}

Consider a target that needs to be covered and a set of agents with coverage capability, $ \mathcal{A}=\left\{a_1,a_2,...,a_N\right\}$, where $N$ represents the total number of agents. Suppose that the strategy of $a_k$ is $\theta_k $, $\theta_k\in \Theta_k$ and $\Theta_k$ is the strategy space of $a_k$.
Given $\theta_k$ and the target, suppose the coverage of $a_k$ with respect to the target is $\mathcal{C}_k$, which is assumed to be a Lebesgue measurable set in the appropriate dimension and is related to $\theta_k$, i.e., $\mathcal{C}_k=\mathcal{C}_k\left(\theta_k\right)$. $\left|\mathcal{C}_k\right|$ represents the Lebesgue measure of $\mathcal{C}_k$.  For example, if $\mathcal{C}_k$ is two-dimensional, it can be regarded as the coverage region.
Let $\theta=\left(\theta_1,...,\theta_N\right)\in\Theta$ be the strategy profile for all agents, where $\Theta=\left(\Theta_1\times\Theta_2\times...\times\Theta_N\right)$ is the combined strategy space of all agents. Let $\theta_{-k}=\left(\theta_1,...,\theta_{k-1},\theta_{k+1},...,\theta_N\right)\in\Theta_{-k}$ be the strategy profile for all agents except $a_k$, where $\Theta_{-k}$ is the value space of $\theta_{-k}$. Let $E_k$ denote the bounded energy penalty function of $a_k$ for implementing a strategy $\theta_k$.  Besides, let $\theta_k=0$ represent that $a_k$ keeps the current state.
\begin{remark}
    Taking the UAV coverage scenario as an example, $\mathcal{C}_k$ is the coverage region of $a_k$ for the target region, which is shown in Fig.~\ref{fig2}. Then, $\left|\mathcal{C}_k\right| $ is the area of the region.  
\end{remark}

Since the performance of the entire system is the main focus, we first give a global performance index to characterize the multi-agent network, which is defined as 
\begin{equation}\label{eqF}
    F\left(\theta\right) = \left| \cup_{k=1}^N \mathcal{C}_k\left(\theta_k\right)  \right|- \gamma \sum_{k=1}^N E_k\left(\theta_k\right),
\end{equation}
where $\gamma$ represents the scale coefficient. Equation (\ref{eqF}) consists of two parts: the coverage performance and the penalty for energy consumption. The combination of two parts enables us to achieve the goal of maximizing coverage objective while saving energy. 

\begin{remark}
    Section~\ref{sec2} and Section~\ref{sec3} focus on achieving the optimum of global objective through distributed decision-making. Therefore, the explicit expression of $F\left(\theta\right)$ is not given in this section. Specific problem instances for UAVs coverage will be given in Section~\ref{sec4}.
\end{remark} 

\subsection{Game-based distributed decision modeling}

\rm In this subsection, the frame of the potential game is introduced to let each agent make decisions in a distributed manner and achieve global near-optimal coverage. Consider the agents as a set of rational players and the OCP can be regarded as a game $G = \left( \mathcal{A},\Theta, \left\{ f_k  \right\}\right)$, where $f_k$ is the local performance objective of $a_k$. In order to enable each agent to make decisions distributively, $f_k$ is designed as 
\begin{equation}\label{eqf}
    f_k\left(\theta_k,\theta_{-k} \right)=\left|\mathcal{C}_k\left(\theta_k\right) - \cup_{l\in \mathcal{N}_k}\mathcal{C}_l\left(\theta_l\right) \right| -\gamma E_k\left(\theta_k\right),
\end{equation}
where the minus sign acting on sets represents the subtraction of these sets and $\mathcal{N}_k$ is the index set of $a_k$'s neighbors. $\mathcal{N}_k$ is defined as
\begin{equation}\label{eqNk}
    \mathcal{N}_k=\left\{ l\in \left\{1,2,...,N\right\} \left|~ l\ne k,~\mathcal{C}_k \cap \mathcal{C}_l \ne \emptyset \right. \right\}.
\end{equation}
Obviously,  $\mathcal{N}_k$ is related to $\theta_k$ and $\theta_{-k}$. Therefore, let 
$$\mathcal{N}_k = \mathcal{N}_k\left(\theta_k,\theta_{-k}\right)$$
represents the index set of $a_k$'s neighbors after agents implement $\theta_k$ and  $\theta_{-k}$.

\begin{remark}
    The first part of \eqref{eqf} suggests that $a_k$ tends to reduce the measure of the coverage that overlaps with its neighbors if it wants to maximize $f_k$. This is a natural index because an agent may not want to cover the target if its neighbors already cover the target. The second part of $f_k$ implies that $a_k$ needs to save energy while maximizing the coverage performance.
\end{remark}

\begin{remark}
      Equation (\ref{eqNk}) means the neighbors of $a_k$ are agents that have coverage that overlaps with $\mathcal{C}_k$. In the UAVs coverage case, the condition that two agents are neighbors, i.e., $ \mathcal{C}_k \cap \mathcal{C}_l \ne \emptyset$, implies that $a_k$ and $a_l$ can simultaneously cover a part of the target region, which is demonstrated in Fig.~\ref{fig2}. The neighbors of $a_2$ are $a_1$ and $a_3$.
\end{remark}

Note that  \eqref{eqf} requires agents to obtain neighborhood information after taking feasible strategies. Thus, the following assumption about the agent's communication capabilities and strategy space is given to guarantee that $a_k$ can calculate its local performance index.

\begin{assumption}\label{assu_nei}
    For all $ k \in \left\{1,2,...,N\right\}$, $a_k$ can obtain the information of $a_l$ for all $l \in \mathcal{N}_k\left(\theta_k, \theta_{-k}\right)$,  $ \theta_k\in\Theta_k$ and  $ \theta_{-k}\in\Theta_{-k}$.
\end{assumption}

Assumption~\ref{assu_nei} places constraints on the agent's strategy space. It requires that the agents will not make decisions beyond their communication capabilities. In other words, $a_k$ can communicate with all agents that may be $a_k$'s neighbors. This not only ensures that the agents can compute  \eqref{eqf}, but also prevents them from adopting overly aggressive strategies that could disrupt the network communication structure.

Then, the game-based distributed OCP and the definition of a powerful tool to model distributed OCP, i.e., the potential games, are given as follows.
\begin{problem}(Game based distributed OCP)\label{gameOCP}
    For each $ k \in \left\{1,2,..., N\right\}$, the agent $a_k$ maximizes (\ref{eqf}) by selecting $\theta_k$  and considering the strategies of its neighbors to maximize the global performance objective (\ref{eqF}), i.e., $a_k$ find $\theta_k$ such that $\theta_k$ and $\theta$ satisfy
    \begin{align*}
     \max_{\theta_k\in\Theta_k}     f_k\left(\theta_k,\theta_{-k} \right)&=\left|\mathcal{C}_k\left(\theta_k\right) - \cup_{l\in \mathcal{N}_k}\mathcal{C}_l\left(\theta_l\right) \right| -\gamma E_k\left(\theta_k\right),\\
     \max_{\theta\in \Theta} F\left(\theta\right) &= \left| \cup_{k=1}^N \mathcal{C}_k\left(\theta_k\right)  \right|- \gamma \sum_{k=1}^N E_k\left(\theta_k\right).
    \end{align*}
\end{problem}

\begin{definition}[Potential game\cite{monderer1996potential}]
A game $G = \left( \mathcal{A}, \Theta  , \left\{ f_k  \right\}\right)$ is called an exact potential
game (or, in short, a potential game) if it admits a global potential function $\phi\left(\theta\right)$, such that
\begin{equation*}
    f_k\left(\tilde{\theta}_k,\theta_{-k}\right)-f_k\left(\theta_k,\theta_{-k}\right)=\phi\left(\tilde{\theta}_k,\theta_{-k}\right)-\phi\left(\theta_k,\theta_{-k}\right)
\end{equation*}
holds for every $k\in\{1,2,...,N\}$ and every $\theta=\left(\theta_k,\theta_{-k}\right)$,  $\tilde{\theta}=\left(\tilde{\theta}_k,\theta_{-k}\right)\in \Theta$.
\end{definition}

In addition, since the strategy space in this paper is continuous, it is difficult to obtain the Nash equilibrium using an iterative distributed algorithm. Thus, the definition of a weaker equilibrium is introduced.
\begin{definition}[$\epsilon$-equilibrium]
For any $\epsilon>0$, a strategy profile $\theta^*=\left(\theta_1^*,...,\theta_N^*\right)$ is an $\epsilon$-equilibrium  if
\begin{equation*}
    f_k\left(\theta_k^*,\theta_{-k}^* \right) \geq f_k\left(\theta_k,\theta_{-k}^*\right)-\epsilon
\end{equation*}
holds for any $\theta_k\in\Theta_k$ and any $k\in\left\{1,2,...,N\right\}$.
\end{definition}

$\epsilon$-equilibrium means any player in a game will not gain a benefit greater than $\epsilon$ from a unilateral change of strategy. If $\epsilon$ tends to 0, the $\epsilon$-equilibrium converges to the Nash equilibrium. Then, the following theorem indicates that Problem~\ref{gameOCP} can be a potential game.

\begin{theorem}[]\label{th1}
Consider the global objective function (\ref{eqF}), local  objective function (\ref{eqf}) and the game $G = \left( \mathcal{A},\Theta, \left\{ f_k  \right\}\right)$.  $G$ is a potential game with potential function $\phi\left(\theta\right)=F\left(\theta\right)$.
\end{theorem}

The reason why we want to establish a potential game is that it has the following advantages.  Firstly, when each player unilaterally and sequentially improves their strategy, the potential function will also be improved. Secondly, if the potential function is bounded, the $\epsilon$-equilibrium must exist, and the equilibrium can be reached within a finite step of improvement (see Lemma 4.1 and Lemma 4.2 in \cite{monderer1996potential}).  Finally, the extreme point of the potential function must equal the equilibrium of the potential game. Based on these features, the next section will design a distributed algorithm to find the $\epsilon$-equilibrium. 

\noindent\textit{\textbf{Proof of Theorem~\ref{th1}}}.
Firstly, some properties of the Lebesgue measure are given. For each $ k,l\in\left\{1,2,...,N\right\}$,
\begin{gather*}
    \left|\mathcal{C}_k\cup \mathcal{C}_l \right|=\left|\mathcal{C}_k\right|+\left|\mathcal{C}_l \right|-\left|\mathcal{C}_k\cap \mathcal{C}_l \right|,\\
    \left|\mathcal{C}_k - \mathcal{C}_l \right| = \left|\mathcal{C}_k\right|-\left|\mathcal{C}_k\cap \mathcal{C}_l \right|
\end{gather*}
holds and further,
\begin{equation*}
    \left|\cup_{k=1}^N \mathcal{C}_k \right|=\sum_{n=1}^N \left(-1\right)^{n+1}\sum_{\stackrel{k_1,...,k_n\in\left\{1,...,N\right\}}{ k_1\ne k_2\ne...\ne k_n}}
    \left|\cap_{l=1}^n \mathcal{C}_{k_l} \right|.
\end{equation*}

On the one hand, according to the properties above, $f_k$ satisfies 
\begin{align*}
    f_k =& \left|\mathcal{C}_k-\cup_{l\in \mathcal{N}_k} \mathcal{C}_l \right|-\gamma E_k\\
    =& \left|\mathcal{C}_k\right|-\left|\mathcal{C}_k\cap\left(\cup_{l\in \mathcal{N}_k} \mathcal{C}_l\right) \right|-\gamma E_k\\
    =& \left|\mathcal{C}_k\right|-\left|\cup_{l\in \mathcal{N}_k}\left( \mathcal{C}_k\cap \mathcal{C}_l\right) \right|-\gamma E_k\\
    =& \left|\mathcal{C}_k\right| -\sum_{n=1}^{N_k} \left(-1\right)^{n+1}\sum_{\stackrel{k_1,...,k_n\in\mathcal{N}_k}{ k_1\ne ...\ne k_n}}
    \left|\mathcal{C}_k\cap\left(\cap_{l=1}^n \mathcal{C}_{k_l} \right)\right|-\gamma E_k,
\end{align*}
where $N_k$ is the number of elements in $\mathcal{N}_k$.
On the other hand, the potential function satisfies
\begin{equation}\label{eqphi1}
    \begin{aligned}
    \phi\left(\theta\right)=&~F\left(\theta\right)\\
    =&\left|\cup_{k=1}^N \mathcal{C}_k\right|-\gamma \sum_{k=1}^N E_k\\
    =&\sum_{k=1}^N\left|\mathcal{C}_k\right| + \sum_{n=2}^N \left(-1\right)^{n+1}
    \sum_{\stackrel{k_1,...,k_n\in\left\{1,...,N\right\}}{ k_1\ne ...\ne k_n}}
    \left|\cap_{l=1}^n \mathcal{C}_{k_l} \right|\\&
    -\gamma \sum_{k=1}^N E_k.
\end{aligned}
\end{equation}

In (\ref{eqphi1}), the terms including $\mathcal{C}_k$ is related to $a_k$. Divide (\ref{eqphi1}) into two parts, which are respectively related to $a_k$ and unrelated to $a_k$. Denote the part unrelated to $a_k$ as $\Gamma_{-k}$.  It yields that
\begin{equation}\label{eqPhi1}
    \begin{aligned}
    \phi\left(\theta\right)=&\left|\mathcal{C}_k\right| + \sum_{n=1}^{N-1} \left(-1\right)^{n}\sum_{\stackrel{k_1,...,k_n\in\left\{1,...,N\right\}}{ k_1\ne ...\ne k_n\ne k}}
    \left|\mathcal{C}_k\cap\left(\cap_{l=1}^n \mathcal{C}_{k_l}\right) \right|\\
    &-\gamma E_k+\Gamma_{-k}\\
    =&\left|\mathcal{C}_k\right|-\sum_{n=1}^{N_k} \left(-1\right)^{n+1}\sum_{\stackrel{k_1,...,k_n\in\mathcal{N}_k}{ k_1\ne ...\ne k_n}}
    \left|\mathcal{C}_k\cap\left(\cap_{l=1}^n \mathcal{C}_{k_l}\right) \right|\\
    &-\gamma E_k+\Gamma_{-k}\\
    =&f_k +\Gamma_{-k},
\end{aligned}
\end{equation}
where the second equal sign is because $\mathcal{C}_k\cap \mathcal{C}_l= \emptyset$ if and only if $l\notin \mathcal{N}_k$ according to (\ref{eqNk}).  Apparently  $\Gamma_{-k}$ will not change when $a_k$ unilaterally change its strategy from $\theta_k$ to $\tilde{\theta}_k$. Therefore,
\begin{equation*}
    \phi\left(\tilde{\theta}_k,\theta_{-k}\right)-\phi\left(\theta_k,\theta_{-k}\right)
    =f_k\left(\tilde{\theta}_k,\theta_{-k}\right)- f_k \left(\theta_k,\theta_{-k}\right).
\end{equation*}
\hfill $\square$

\section{Distributed optimal coverage searching algorithm design with convergence guarantee}\label{sec3}

In this section, an algorithm to find the equilibrium of the game in Problem \ref{gameOCP}  is designed and analyzed.

When $a_k$  unilaterally improves its strategy from $\theta_k$ to $\tilde{\theta}_k$, define the change of local  objective   function as \textit{regret value}, $R_k$, i.e.,
$$R_k\left(\tilde{\theta}_k,\theta_k,\theta_{-k}\right) = f_k\left(\tilde{\theta}_k,\theta_{-k}\right)-f_k\left(\theta_k,\theta_{-k}\right).$$
Inspired by \cite{sun2021game}, the definition of \textit{$\epsilon$-innovator} is first proposed to enable agents to improve the global potential function by updating strategies based on local information.

\begin{definition}[$\epsilon$-innovator]\label{innnovator}
    For any $ \epsilon>0$, an agent $a_k$ is called an $\epsilon$-innovator in a strategy iteration if it satisfies the following properties.
    
    1) $a_k$ has the largest regret value in all neighbors after executing the strategy, and the regret value is greater than $\epsilon$, i.e., 
        \begin{gather*}
        \begin{cases}
              ~R_k\left(\tilde{\theta}_k,\theta_k,\theta_{-k}\right)\geq R_l\left(\tilde{\theta}_l,\theta_l,\theta_{-l}\right),\forall l\in \mathcal{N}_k\left(\tilde{\theta}_k,\theta_{-k}\right),\\
            ~R_k\left(\tilde{\theta}_k,\theta_k,\theta_{-k}\right)>\epsilon,
        \end{cases} 
        \end{gather*}
        
        2) $a_k$ has the smallest index among all neighbors with the same regret value, i.e., if $\exists l \in\mathcal{N}_k\left(\tilde{\theta}_k,\theta_{-k}\right)$, s.t. $R_k=R_l$, then $k<l$.
\end{definition}

\begin{figure*}[!htbp]
\centering
\includegraphics[width=0.8\textwidth]{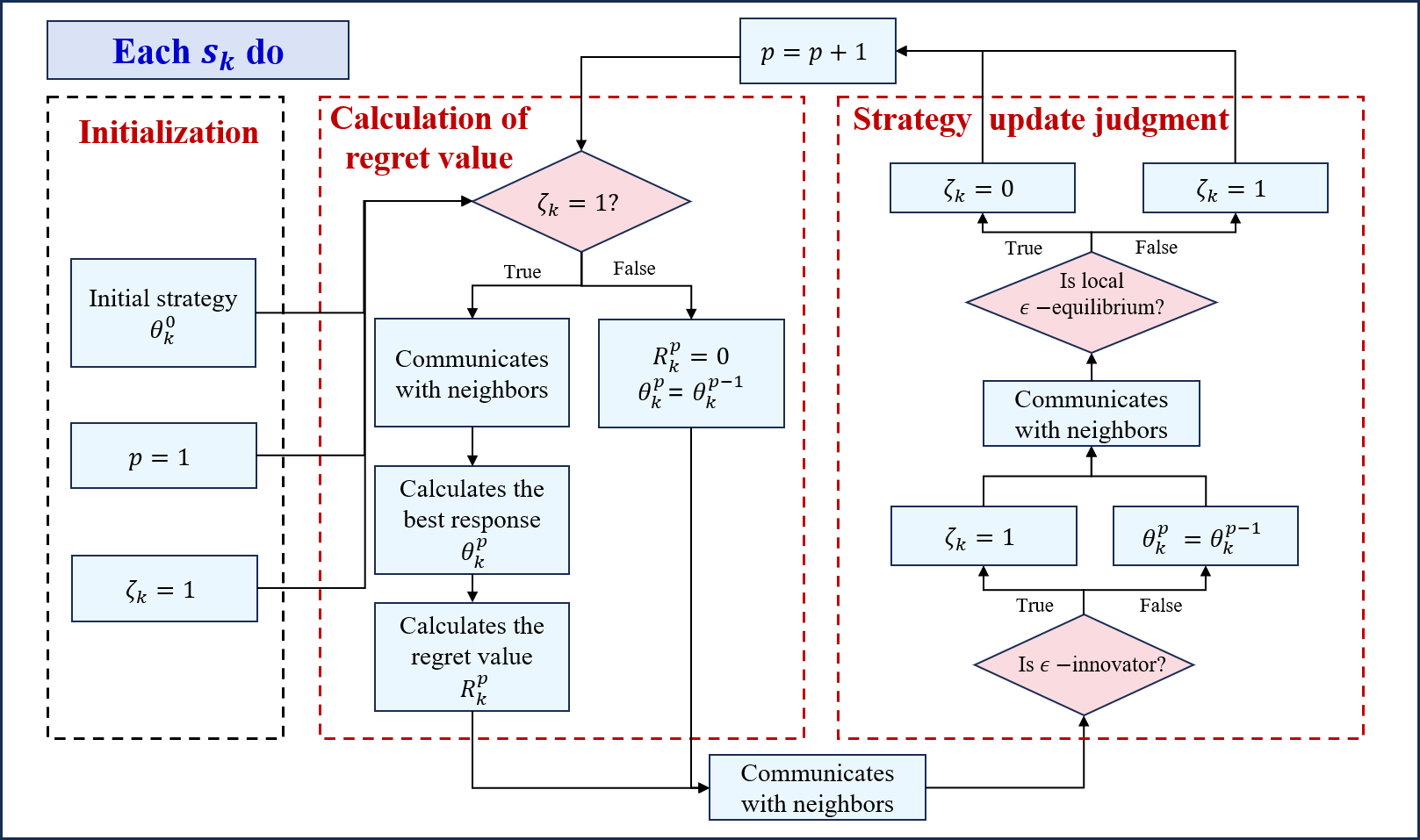}
\caption{Flowchart of Distributed Optimal Coverage Searching Algorithm.}
\label{figflow}
\end{figure*}

By Definition \ref{innnovator}, each neighborhood permits at most one $\epsilon$-innovator. This implies that any two concurrent $\epsilon$-innovators must be mutually non-adjacent. Then, a local generalization of $\epsilon$-equilibrium is provided to help agents determine whether they need to calculate and update in the next round of strategy iteration.
\begin{definition}[local $\epsilon$-equilibrium]
    For any $\epsilon>0$, any $a_k\in \mathcal{A}$ and corresponding index set of neighbors  $\mathcal{N}_k\left(\theta_k,\theta_{-k}\right)$,  $a_k$ is said to reach a local $\epsilon$-equilibrium in a policy iteration if it satisfies
    $$\forall ~ l \in  \mathcal{N}_k\bigcup\{k\},~R_l\left(\tilde{\theta}_l,\theta_l,\theta_{-l}\right)\leq \epsilon.$$
    
\end{definition}
\begin{remark}
    Local $\epsilon$-equilibrium indicates that all agents in the neighborhood of $a_k$ cannot benefit more than $\epsilon$ by unilaterally updating their strategies and thus $a_k$ will not change strategy in the next iteration. 
\end{remark}
Then, Algorithm~\ref{DOCS} is developed to solve the distributed OCP, and its flowchart is illustrated in Fig.~\ref{figflow}. 
In Algorithm \ref{DOCS}, the auxiliary variable $\zeta_k$ is used to reduce computation cost by determining whether $a_k$ calculates its best response and regret value, i.e., line 5 and line 6 in Algorithm~\ref{DOCS}. These steps are the most computationally demanding part of each iteration. After calculating and exchanging the regret value, only the $\epsilon$-innovators can update their strategies, and each agent checks whether they have reached a local $\epsilon$-equilibrium. If it's false, it indicates that some agents may be $\epsilon$-innovators in the neighborhood of $a_k$ in the current iteration, which means the best response of $a_k$ may change. Consequently, set $\zeta_k=1$, and $a_k$ can update its strategy in the next iteration. Otherwise, set $\zeta_k=0$.

\begin{algorithm}[!ht]
\footnotesize
\caption{Distributed Optimal Coverage Searching (DOCS)}
\label{DOCS}
\begin{algorithmic}[1]
        \REQUIRE Agents set $S$, convergence accuracy $\epsilon$, total number of iterations $P$;
    \ENSURE Output the $\epsilon-$equilibrium;
    \STATE Initialize strategies $\theta_k^0$ and auxiliary variable $\zeta_k=1$, $k=1,2,...,N$;
    \FOR{$p=1,2,...,P$}
     \FOR{$k=1,2,...,N$, $a_k$ synchronously}
     \IF{$\zeta_k=1$,}
    \STATE Calculate $\theta_k^p=\underset{\theta_k\in \Theta_k}{\text{argmax}}f_k\left( \theta_k,\theta_{-k}^{p-1} \right)$;
    \STATE Calculate $R_k^p=R_k\left(\theta_k^p, \theta_k^{p-1},\theta_{-k}^{p-1} \right)$;
    \ELSE
    \STATE Set $R_k^{p}=0$, $\theta_k^p = \theta_k^{p-1}$;
    \ENDIF
     \STATE Communicate with neighbors;
    \IF{$a_k$ is an $\epsilon$-innovator,}
    \STATE Set $\zeta_k=1$;
    \ELSE
    \STATE $\theta_k^p=\theta_k^{p-1}$;
    \ENDIF
    \STATE Communicate with neighbors and update regret value:
        Calculate $R_k^p=R_k\left(\theta_k^p, \theta_k^{p-1},\theta_{-k}^{p-1} \right)$;
    \IF{$a_k$ reachs the local $\epsilon$-equilibrium}
        \STATE Set $\zeta_k=0$;
        \ELSE 
        \STATE Set $\zeta_k=1$ ;
        \ENDIF
    \ENDFOR
    \ENDFOR
\end{algorithmic}
\end{algorithm}

    In Algorithm \ref{DOCS}, the selection of $P$ should depend on the convergence accuracy $\epsilon$ and the prior knowledge about problems (see Theorem~\ref{th2}). Besides, even if $P$ is chosen much larger than the number of iterations required for convergence, the extra iterations will also proceed significantly faster because of the local $\epsilon$-equilibrium judgment mechanism (lines 17 to 21 in Algorithm \ref{DOCS}). Please see Fig.~\ref{DOCS-DT2A} for illustration.

\begin{remark}
    Unlike traditional distributed approaches like SAP\cite{sun2017time} and BRR\cite{ai2008optimality}, which employ centralized agent selection for strategy updates, making them pseudo-distributed, Algorithm~\ref{DOCS} leverages the $\epsilon$-innovator mechanism to achieve genuine distributed operation.
\end{remark}
\begin{remark}
    Compared to DT2A in \cite{sun2021game}, Algorithm~\ref{DOCS} has two key innovations. Firstly, it extends the innovator mechanism that is only applicable to discrete strategy space to the $\epsilon$-innovator mechanism that applies to continuous strategy space scenarios. Secondly, it proposes a local $\epsilon$-equilibrium judgment mechanism to reduce the computational complexity.
\end{remark}

Consider the correctness of $\epsilon$-innovators obtained by Algorithm \ref{DOCS}. If we delete lines 17 to 21 in Algorithm~\ref{DOCS} and let $\zeta_k=1$ hold for every iteration, then Algorithm~\ref{DOCS} can decide the correct $\epsilon$-innovators because it calculates $R_k^p$ just as the definition of regret value (line 6 in Algorithm~\ref{DOCS}). After we execute lines 17 to 21 in Algorithm~\ref{DOCS}, there is only one possible case that causes Algorithm~\ref{DOCS} to identify a wrong $\epsilon$-innovator $a_k$. That case is $\zeta_k=0$, while $R_k\left(\theta_k^p, \theta_k^{p-1},\theta_{-k}^{p-1} \right)>\epsilon$, where $\theta_k^p=\underset{\theta_k\in \Theta_k}{\text{argmax}}f_k\left( \theta_k,\theta_{-k}^{p-1} \right)$. In other words, the real regret value of $a_k$ is greater than $\epsilon$ (which means $a_k$ can be an $\epsilon$-innovator) in an iteration, but Algorithm~\ref{DOCS} lets $R_k^p=0$. The following lemma indicates that this situation will not occur.
\begin{lemma}\label{lemma2}
Algorithm \ref{DOCS} can identify the correct $\epsilon$-innovators in each iteration.
\end{lemma}

Then,  Lemma~\ref{lemma1} is given for analyzing the convergence of Algorithm \ref{DOCS}. The proofs of Lemma~\ref{lemma2} and Lemma~\ref{lemma1} are provided in Appendix~\ref{appen3} and Appendix~\ref{appen1}.

\begin{lemma}[]\label{lemma1}
Consider the exact potential game $G$ and Algorithm \ref{DOCS}. For any $ p \in\left\{ 1,2,..., P-1\right\}$,  it holds that
\begin{equation*}
    \phi\left(\theta^{p+1}\right)-\phi\left(\theta^p\right)= \sum_{k\in\mathcal{I}^{p+1}} R_k^{p+1},
\end{equation*}
where  $\theta^p$ is the strategy profile obtained after  iterations $p$ of Algorithm \ref{DOCS} and $\mathcal{I}^{p+1}$ is the index set of $\epsilon$-innovators in iteration $\left(p+1\right)$, that is
\begin{align*}
    \mathcal{I}^{p+1}\triangleq &\left\{k\left|k\in \left\{1,2,...,N\right\}{\rm and} ~a_k~ {\rm is}~ \epsilon\text{-}{\rm innovator}\right.\right.\\      
      &\qquad\qquad\qquad \left. {\rm~ in~ the}~ (p+1){\rm \text{-}th~ iteration}\right\}.
\end{align*}
\end{lemma}

Lemma~\ref{lemma1} implies that in the sense of $ \phi\left(\theta\right)$'s value, all $\epsilon$-innovators (for example, $\kappa$ $\epsilon$-innovators) update strategies simultaneously in one iteration is equivalent to that they update strategies unilaterally and sequentially in $\kappa$ iterations. This is because the $\epsilon$-innovators are not neighbors of each other and their strategic updates will not affect each other in one iteration. Therefore, the $\epsilon$-innovators mechanism can enhance convergence speed, achieved by permitting simultaneous strategy updates across multiple agents per iteration. Then, the convergence of Algorithm~\ref{DOCS} is proven under a weak assumption.
\begin{assumption}\label{assumption1}
    For any $ \theta \in \Theta$,  there exists two constants, $ \phi_{min} $ and $\phi_{max}$, such that $\phi_{min}\leq \phi\left(\theta\right)\leq \phi_{max}$.
\end{assumption}

\begin{theorem}\label{th2}
    Consider Problem~\ref{gameOCP} and Algorithm~\ref{DOCS} under Assumption~\ref{assumption1}. For any $ \epsilon>0$, Algorithm~\ref{DOCS} can output an $\epsilon$-equilibrium after the iterations of  
     \begin{equation}\label{eqP}
        P=\lfloor\frac{\phi_{max}-\phi_{min}}{\epsilon}\rfloor+1,
    \end{equation}
     where $\lfloor \cdot\rfloor$ is the floor bracket.
\end{theorem}

The proof of Theorem~\ref{th2} is provided in Appendix~\ref{appen2}.
\begin{remark}
    In many practice scenarios, the lower bound and the upper bound of $\phi\left(\theta\right)$ can be easily obtained. For the UAVs coverage case, the upper bound can be chosen as the total area of the target region, and the lower bound can be selected as the value of $\phi\left(\theta\right)$ before optimization.
\end{remark}
Theorem~\ref{th2} suggests that if the upper bound and lower bound of $\phi\left(\theta\right)$ can be obtained in advance, then the selection of $P$ can be determined and the convergence of Algorithm~\ref{DOCS} can be guaranteed. 
Besides, the output of Algorithm~\ref{DOCS} is $\epsilon$-equilibrium of the potential game, i.e., the extremum of the potential function $\phi(\theta)$ with the tolerance error $\epsilon$, and is not the global maximum of $\phi(\theta)$. Actually, the maximum value is challenging to obtain even if using the centralized algorithm when the performance index is complicated. Therefore, the focus of this paper is not on how to escape the local extremum and search for the global maximum. Some random techniques can help find a better extremum (for example, see \cite{gao2022coverage,sun2021game}), but they will also increase the computation time of the algorithm.

\section{Application to UAV network and simulation study}
Consider the coverage optimization scenario of the UAVs on the two-dimensional target region as shown in Fig.~\ref{fig2}. The coverage region of each drone is a circular region centered on its projection point on the plane. Through local information exchange, each drone tries to make decisions to maximize the overall coverage area of the target region while minimizing the energy consumption required for movement.

\subsection{Optimal coverage problem in UAV network}\label{sec4}

Consider a set of UAV $\mathcal{A}=\left\{a_1,a_2,...,a_N\right\}$ on two-dimensional plane. The strategy of $a_k$ is $\theta_k = \left[\theta_{k_x},\theta_{k_y}\right]^{\rm T}$, which represents the displacement from the initial deployment location to the final deployment location. The feasible deployment location for $a_k$ is a rectangular region near the initial position, i.e.,  $\theta_k$ satisfies
$$\theta_k\in \Theta_k \triangleq \left\{\left[\theta_{k_x},\theta_{k_y}\right]^{\rm T}\left| \left| \theta_{k_x}\right| \leq L_{k_x}, ~\left| \theta_{k_y}\right| \leq L_{k_y}\right.\right\},$$
where  $ L_{kx}$ and $ L_{ky}$ are known parameters. Let $[x_k ,y_k]^{\rm T}$ represent the initial position of $a_k$ and let $\rho_k$ be the cover radius of $a_k$. Then, a point $[x,y]^{\rm T}$ is coverable for $a_k$ at the initial position if it satisfies 
\begin{equation*}
    \sqrt{ \left(x-x_k\right)^2 + \left(y-y_k\right)^2} \leq \rho_k.
\end{equation*}

Let $\mathcal{C}_k$ be the coverage set of $a_k$ for the target region and $\left|\mathcal{C}_k\right|$ is the coverage area of $a_k$. In this scenario, $\mathcal{C}_k$ will be a function of $\theta_k$. Specifically, denote
\begin{align*}
   & \tau_k(x,y,\theta_k)\\
   &= \begin{cases}
        1,& {\rm if}~ \sqrt{ \left(x-\left(x_k+\theta_{k_x}\right)\right)^2 + \left(y-\left(y_k+\theta_{k_y}\right)\right)^2} \leq \rho_k,\\
        0, & {\rm otherwise.}
    \end{cases}
\end{align*}
as the coverage function of $a_k$. $\tau_k(x,y,\theta_k)=1$ if $a_k$ can cover $[x,y]^{\rm T}$ after it excutes strategy $\theta_k$. Otherwise, it equals 0. Denote the target region as $\mathcal{S}_{\rm tar}$.  We have that 
\begin{equation*}
    \left|\mathcal{C}_k\left(\theta_k\right)\right|= \int_{\mathcal{S}_{\rm tar}}\tau_k(x,y,\theta_k){\rm d} x {\rm d} y.
\end{equation*}
Define the peak-shaving function
\begin{equation*}
    g\left(x\right)=
    \begin{cases}
        1,~\text{if}~ x\geq1,\\
        x,~\text{if}~ x<1.
    \end{cases}
\end{equation*}
Then, the global coverage performance can be obtained by 
\begin{equation*}
    \left|\bigcup_k \mathcal{C}_k\left(\theta_k\right) \right|=\int_{\mathcal{S}_{\rm tar}}g\left(\sum_k\tau_k(x,y,\theta_k)\right){\rm d} x {\rm d} y,
\end{equation*}
where $g\left(\sum_k\tau_k(x,y,\theta_k)\right)$ satisfies that it equals 1 if at least one UAV can cover $[x,y]^{\rm T}$. Otherwise, it equals 0. 
For the local performance index of $a_k$, one can obtain that
\begin{equation*}
\begin{aligned}
    &\left|\mathcal{C}_k\left(\theta_k\right) - \bigcup_{l\in\mathcal{N}_k} \mathcal{C}_l\left(\theta_l\right) \right|\\
    &=\int_{\mathcal{S}_{\rm tar}} \tau_k\left(x,y,\theta_k\right) \left(1-g\left(\sum_{l\in\mathcal{N}_k}\tau_l\left(x,y,\theta_l\right)\right)\right) {\rm d} x {\rm d} y,
\end{aligned}
\end{equation*}
where $\tau_k\left(x,y,\theta_k\right) \left(1-g\left(\sum_{l\in\mathcal{N}_k}\tau_l\left(x,y,\theta_l\right) \right)\right)$ satisfies that it equals 1 if  $a_k $ can cover $ [x,y]^{\rm T}$ and all of the neighbors of $a_k $ can not cover $ [x,y]^{\rm T}$. Otherwise, it equals 0. 

The energy consumption function is defined as    
\begin{equation*}
  E_k\left(\theta_k \right)=\theta_k^2,
\end{equation*}

Finally, the explicit expressions of $f_k$ and $F$ are
\begin{equation}\label{eqfsa}
\begin{aligned}
    &f_k\left(\theta_k,\theta_{-k} \right)\\=&\int_{\mathcal{S}_{\rm tar}} \tau_k\left(x,y,\theta_k\right) \left(1-g\left(\sum_{l\in\mathcal{N}_k}\tau_l\left(x,y,\theta_l\right)\right)\right) {\rm d} x {\rm d} y \\
    &- \gamma \theta_k^2,
\end{aligned}
\end{equation}
\begin{equation}\label{eqFsa}
    F\left(\theta\right) = \int_{\mathcal{S}_{\rm tar}}g\left(\sum_{k=1}^N\tau_k(x,y,\theta_k)\right){\rm d} x {\rm d} y- \gamma \sum_{k=1}^N \theta_k^2,
\end{equation}

Then, the multi-agent distributed OCPs in the UAV network are as follows.
\begin{problem}[Game based distributed OCP in the UAV network]\label{distributedOCP}
    For each $ k \in \left\{1,2,..., N\right\}$, $a_k$ selects $\theta_k$ to maximize (\ref{eqfsa}) with the information from its neighbors such that the UAV network can achieve the extremum of (\ref{eqFsa}), i.e., each $a_k$ find $\theta_k$ such that $\theta_k$ and $\theta$ satisfy 
    \begin{align*}
        \max_{\theta_k\in\Theta_k}& f_k\left(\theta_k,\theta_{-k} \right)\\=&\int_{\mathcal{S}_{\rm tar}} \tau_k\left(x,y,\theta_k\right) \left(1-g\left(\sum_{l\in\mathcal{N}_k}\tau_l\left(x,y,\theta_l\right)\right)\right) {\rm d} x {\rm d} y \\
    &- \gamma \theta_k^2,\\
 \max_{\theta\in\Theta}&F\left(\theta\right) = \int_{\mathcal{S}_{\rm tar}}g\left(\sum_{k=1}^N\tau_k(x,y,\theta_k)\right){\rm d} x {\rm d} y- \gamma \sum_{k=1}^N \theta_k^2.
    \end{align*}
\end{problem}

\subsection{Simulation study}\label{sec5}
In this subsection, the proposed algorithm, DOCS, is simulated under several scenarios and compared with other typical distributed algorithms from prior work on potential-game-based OCPs. The solver for the best response step in the algorithm is a MATLAB solver, patternsearch. Consider an initial scenario in which several UAVs are randomly scattered in a two-dimensional plane. The target region satisfies
\begin{equation*}
    \mathcal{S}_{\rm tar}=\left\{ [x,y]^{\rm T}\left|
    \begin{matrix}
        \underline{x}_1\leq x \leq \bar{x}_1 ~{\rm or}~ \underline{x}_2\leq x \leq \bar{x}_2,\\
        \underline{y}_1\leq y \leq \bar{y}_1  ~{\rm or}~ \underline{y}_2\leq y \leq \bar{y}_2
    \end{matrix}\right. \right\}.
\end{equation*} 
The parameters used in this section are listed in Table~\ref{tab2}.

\begin{table}[htbp]
\renewcommand{\arraystretch}{1.5}
\caption{Parameters setup\label{tab2}}
\centering
\begin{tabular}{cc}
\hline
Parameter & Value\\
\hline
Convergence accuracy  & $\epsilon=2$\\
Total number of iterations  & $P=40$ \\
Scale coefficient  & $\gamma=0.2$ \\
Bounds of the target region & \makecell[l]{ $\left[\underline{x}_1,\bar{x}_1, \underline{x}_2,\bar{x}_2\right]=\left[0, 200, 300, 600 \right]$ m\\
$\left[\underline{y}_1,\bar{y}_1, \underline{y}_2,\bar{y}_2\right]=\left[0, 200, 400, 600 \right]$  m  }\\
Coverage radius & $\rho_k = 60$ m \\
Bounds of $\theta_k$  & $L_{k_x}=L_{k_y}=60$ m\\
\hline
\end{tabular}
\end{table}

\begin{figure}[!htbp]
\centering
\includegraphics[width=0.45\textwidth]{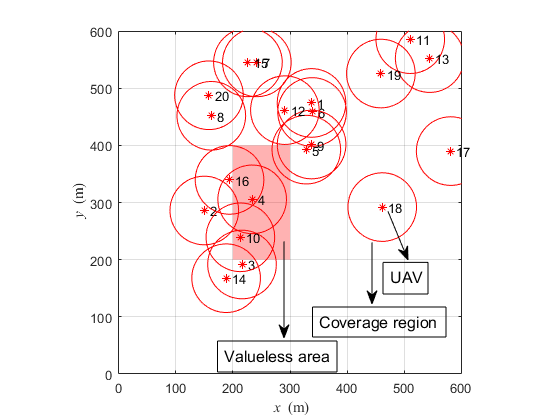}
\caption{Illustration of the initial UAV coverage scenario.}
\label{fig4}
\end{figure}

Fig.~\ref{fig4} illustrates the initial scenario with 20 UAVs according to the parameters in Table~\ref{tab2}. In Fig.~\ref{fig4}, the range of the coordinate axis is the target region, except for the red area. The asterisks represent the initial positions of the UAVs, and the circles centered on the asterisks represent the corresponding coverage areas of the UAVs. Fig.~\ref{fig4} indicates that there are large overlapping coverage areas of the initial UAV network. 

\subsubsection{Simulation results under different scenarios}
\begin{figure}[!htbp]
\centering
\includegraphics[width=0.45\textwidth]{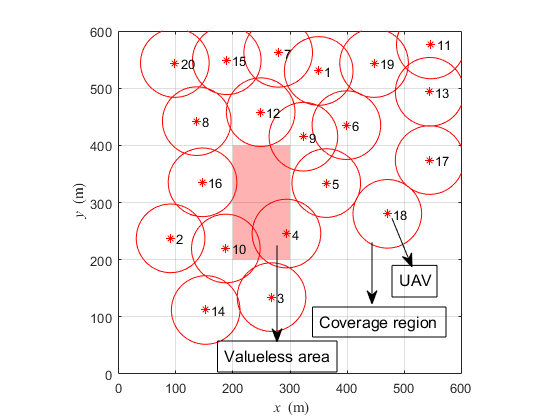}
\caption{The final UAV coverage scenario obtained by DOCS with a valueless area.}
\label{ocp_sce20}
\end{figure}

\begin{figure}[!htbp]
\centering
\includegraphics[width=0.45\textwidth]{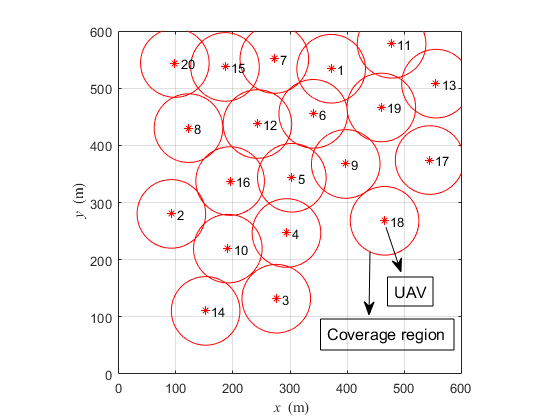}
\caption{The final UAV coverage scenario obtained by DOCS without a valueless area.}
\label{ocp_sce20_2}
\end{figure}

This subsection simulates distinct scenarios.  After DOCS optimization, Fig.~\ref{ocp_sce20} shows the final positions of 20 UAVs corresponding to the initial scenario in Fig.~\ref{fig4}. It can be obtained that after being optimized by DOCS, the UAV network is dispersed to increase the total coverage area and reduce overlapping coverage. Note that the UAVs tend not to cover the red area in Fig.~\ref{fig4} since this area has no coverage value. As a comparison, Fig.~\ref{ocp_sce20_2} demonstrates the positions result under the scenario where the red region has the same value as the other region.

\begin{figure}[!htbp]
\centering
\includegraphics[width=0.45\textwidth]{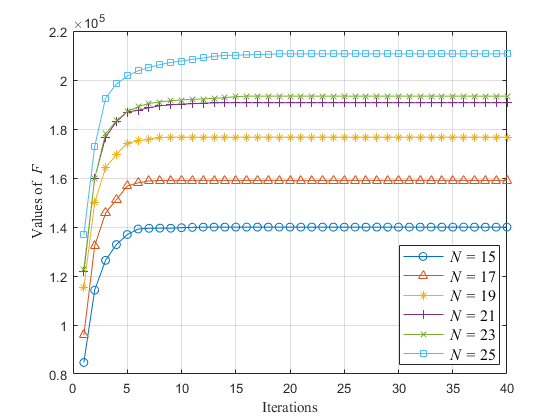}
\caption{The values of global performance obtained by DOCS for different numbers of UAVs.}
\label{t-val_DOCS}
\end{figure}

\begin{figure}[!htbp]
\centering
\includegraphics[width=0.45\textwidth]{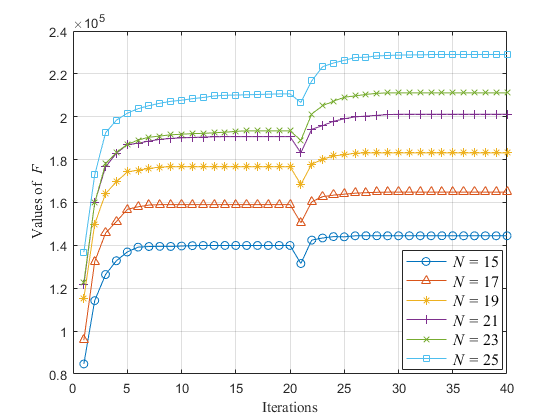}
\caption{The values of global performance obtained by DOCS for different numbers of UAVs under a re-optimization scenario when a UAV is damaged.}
\label{damage_sce}
\end{figure}

Fig.~\ref{t-val_DOCS} shows the values of global performance optimized by DOCS.
Fig.~\ref{t-val_DOCS} indicates that the proposed algorithm can rapidly converge to the final solution. Additionally, consider a re-optimization scenario where a UAV breakdown occurs after the first optimization, causing the network to begin optimizing again. The values of global performance are shown in Fig~\ref{damage_sce}. It can be seen from Fig.~\ref {damage_sce} that although the performance decreases due to the sudden damage, the remaining UAVs compensate for the performance loss by optimizing and maneuvering again.

\subsubsection{Simulation results compared with other methods}
In this part, two distributed methods for potential-game-based OCPs are compared with the proposed algorithm, DOCS. These methods are described as follows.

(1) BRR (Best Response Rule \cite{ai2008optimality}): A distributed stochastic algorithm where one agent is randomly selected per iteration to update its strategy to the best response, while others keep their current strategy.

(2) $\epsilon$-DT2A: A distributed approach adjusted from DT2A\cite{sun2021game} for the potential game of satellite task allocation with an innovator mechanism. In $\epsilon$-DT2A, the innovator in DT2A is replaced by $\epsilon$-innovator to adjust the strategy space from discrete to continuous. DT2A also has a memory mechanism. In $\epsilon$-DT2A, the memory length is set to 1 for direct comparison with DOCS. The key difference in this setting is that \textit{all} agents in $\epsilon$-DT2A calculate their best response each iteration, whereas in DOCS, only agents with $\zeta_k=1$ do so. 

\begin{figure}[!htbp]
\centering
\includegraphics[width=0.45\textwidth]{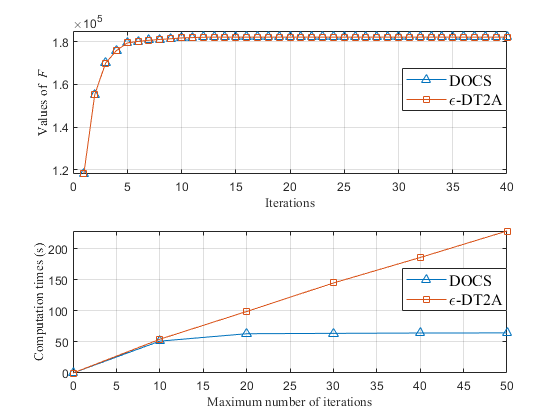}
\caption{The values of global performance and computation times of DOCS and $\epsilon$-DT2A.}
\label{DOCS-DT2A}
\end{figure}

\begin{figure}[!htbp]
\centering
\includegraphics[width=0.45\textwidth]{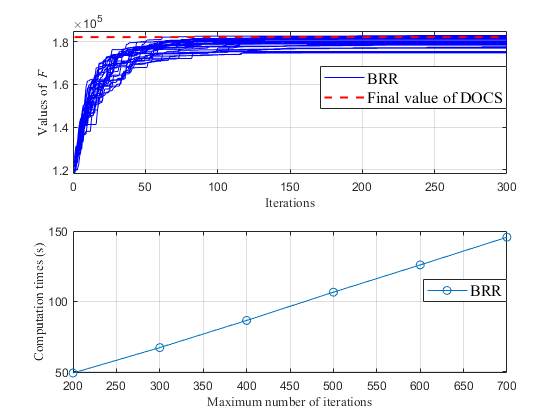}
\caption{The values of global performance and computation times of BRR.}
\label{BRR}
\end{figure}

The results of 30 simulation runs for each method with 20 UAVs are summarized in Table~\ref{tab5}, Fig.~\ref{DOCS-DT2A} and Fig.~\ref {BRR}. Table~\ref{tab5} gives the values of the global performance and computation times for different methods. It indicates that BRR achieves the best global performance values, but the results are affected by randomness. DOCS and $\epsilon$-DT2A achieve global performance values similar to BRR's best value, with DOCS exhibiting the shortest computation time. Note that both DOCS and $\epsilon$-DT2A are deterministic algorithm, hence the global performance values are unchanged.

\begin{table}[!htbp]
\renewcommand{\arraystretch}{1.5}
\caption{Values of the global performance and computation times for different methods under 30 simulations\label{tab5}}
\centering
\begin{tabular}{ccccccc}
\hline
\multirow{2}{*}{Method} & \multicolumn{3}{c}{Value ($10^5$)} & Computation time (s)\\
                        & Average  & Best  & Worst  & Average       \\ \hline
BRR                     & $1.797$    & \bf 1.829  & $1.747$        &    67.3             \\
$\epsilon$-DT2A                    &  $1.821$     &  $1.821$  &  $1.821$   & 145.1         \\
DOCS                    & \bf 1.821    &  $1.821$  & \bf 1.821   & \bf 63.3      \\
\hline
\end{tabular}
\end{table}

Fig.~\ref{DOCS-DT2A} compares the values of the global performance and computation times of DOCS and $\epsilon$-DT2A. The curves of $F$ versus iteration for both methods coincide perfectly.  This confirms that the local $\epsilon$-equilibrium judgment mechanism proposed in DOCS can correctly identify $\epsilon$-innovators. Fig.~\ref{DOCS-DT2A} also shows that although both DOCS and $\epsilon$-DT2A converge within 15 iterations, the extra iterations of DOCS require significantly less time than those of $\epsilon$-DT2A.

Fig.~\ref{BRR} demonstrates the computational inefficiency of BRR across 30 simulation runs. The results indicate that BRR requires over 100 iterations to converge and fails to match the performance of DOCS in most cases. Additionally, its computation time exhibits a continuous increase as the maximum iteration count grows.

\begin{remark}
BRR and DT2A lack theoretical upper bounds on the iteration number required for convergence. In practice, this necessitates setting a potentially overly conservative maximum number of iterations. In contrast, Theorem~\ref{th2} provides an upper bound guaranteeing the convergence of the DOCS algorithm.
\end{remark}

In summary, the proposed DOCS algorithm achieves superior global performance and computational efficiency, offering a marked advantage over traditional distributed methods for potential-game-based OCPs.

\section{Conclusion}\label{sec6}
In this paper, a distributed decision-making method for multi-agent optimal coverage problems (OCPs) in the framework of game theory is studied. In particular, a global performance objective considering both coverage performance and energy consumption is given. In order to solve the OCP, we divide the global performance objective into local performance objectives corresponding to each agent and establish a multi-agent game based distributed OCP. The equivalence between the equilibrium of the game and the extreme of the global performance objective is obtained by proving that the game is a potential game. Then, a distributed optimal coverage searching algorithm is designed to find the equilibrium solution of the multi-agent game, and the convergence within a finite number of iterations is strictly proved. Finally, the proposed method is applied to the UAV network scenarios to maximize the total coverage area for a target region while saving energy. The simulation results indicate that, compared to the typical distributed algorithms in OCP, the proposed distributed method converges rapidly while having close optimization results, which shows the efficiency and validity of our method.


\begin{appendices}
\section{Proof of Lemma \ref{lemma2}}\label{appen3}
Lemma~\ref{lemma2} indicates that Algorithm~\ref{DOCS} can correctly identify whether $a_k$ is an $\epsilon$-innovator  at each iteration even if $\zeta_k=0$, $k\in \{1,...,N\}$. We prove Lemma~\ref{lemma2} by induction and contradiction. 

\textbf{Base case ($p=1$):} In the first iteration, $\zeta_k = 1$ for all $k \in \{1, \ldots, N\}$ (initialization, Algorithm~\ref{DOCS}). When $\zeta_k = 1$, Algorithm~\ref{DOCS} explicitly calculates $R^p_k$ according to the definition of the regret value (line 6). Therefore, Algorithm~\ref{DOCS} correctly identifies $\epsilon$-innovators in the first iteration.

\textbf{Inductive hypothesis ($1\leq p \leq p_1-1$):} Assume Lemma~\ref{lemma2} holds in iterations $p$, $\forall p \in \{1, 2, ..., p_1 - 1\}$, where $p_1 > 1$. That is, Algorithm~\ref{DOCS} correctly identified all $\epsilon$-innovators  in iterations $1$ through iterations $p_1 - 1$.

\textbf{Inductive step ($p = p_1$):} Now prove by contradiction that Lemma~\ref{lemma2} holds for iteration $p_1$. Suppose that there exists some agent $a_{k_1}$, $k_1 \in \{1, ..., N\}$, such that:

\begin{enumerate}
    \item $\zeta_{k_1} = 0$ (Algorithm~\ref{DOCS} did not flag $a_{k_1}$ for best response calculation).\\
    \item The actual regret value of $a_{k_1}$ exceeds $\epsilon$, i.e., $ R_{k_1}\left(\theta_{k_1}^{p_1}, \theta_{k_1}^{{p_1}-1},\theta_{-{k_1}}^{{p_1}-1} \right)>\epsilon$, where
    $$\theta^{p_1}_{k_1} = \underset{\theta_{k_1} \in \Theta_{k_1}}{\operatorname{argmax}} f_{k_1}\left(\theta_{k_1}, \theta^{p_1-1}_{-k_1}\right).$$
\end{enumerate}
Because $\zeta_{k_1} = 0$ in iteration $p_1$, it follows from Algorithm~\ref{DOCS} (lines 17-21) that 
\begin{equation}\label{eqRlp1}
    R_l^{p_1-1}\leq\epsilon,~\forall l \in  \mathcal{N}_{k_1}\left(\theta_{k_1}^{p_1-1},\theta_{-{k_1}}^{{p_1}-1}\right)\bigcup\{k_1\}.
\end{equation}
By the inductive hypothesis, the $\epsilon$-innovators identified in iteration $p_1 - 1$ were correct. Equation \eqref{eqRlp1} signifies that in iteration $p_1 - 1$, the neighborhood of $a_{k_1}$ (including $a_{k_1}$ itself) had achieved the local $\epsilon$-equilibrium. This means no agent in this neighborhood (including $a_{k_1}$) could improve its local objective by more than $\epsilon$ through a unilateral strategy change. Consequently, $ R_{k_1}\left(\theta_{k_1}^{p_1}, \theta_{k_1}^{{p_1}-1},\theta_{-{k_1}}^{{p_1}-1} \right)  \leq \epsilon$. This contradicts our initial assumption $ R_{k_1}\left(\theta_{k_1}^{p_1}, \theta_{k_1}^{{p_1}-1},\theta_{-{k_1}}^{{p_1}-1} \right)>\epsilon$.

This contradiction implies our initial supposition must be false. Therefore, no such $k_1$ exists where $\zeta_{k_1}=0$ and the actual regret value is larger than $\epsilon$. Algorithm~\ref{DOCS} correctly identifies all $\epsilon$-innovators in iteration $p_1$.
By the principle of mathematical induction, Lemma~\ref{lemma2} holds for all iterations $p \in \{1, 2, \ldots, P\}$. 
    \hfill $\square$

\section{Proof of Lemma \ref{lemma1}}\label{appen1}

According to Lemma~\ref{lemma2}, Algorithm~\ref{DOCS} can generate correct innovators. Suppose $a_k$ and $a_l$ are the only two $\epsilon$-innovators in the $(p+1)$-th iteration.  According to the Definition \ref{innnovator}, it can be obtained that $a_k$ and $a_l$ are not neighbors of each other, $R_k^{p+1}>\epsilon$ and $R_l^{p+1}>\epsilon$. As a result, the change of $\theta_l^{p+1}$ has no effect on $f_k \left(\theta_k^{p+1},\theta_{-k}^p\right)$ and the change of $\theta_k^{p+1}$ has no effect on $f_l \left(\theta_l^{p+1},\theta_{-l}^p\right)$, i.e.,
    \begin{gather}
        f_k \left(\theta_k^{p+1},\theta_{-k}^{p+1}\right)=f_k \left(\theta_k^{p+1},\theta_{-k}^p\right)\label{eqfk},\\
        f_l \left(\theta_l^{p+1},\theta_{-l}^{p+1}\right)=f_l \left(\theta_l^{p+1},\theta_{-l}^p\right).\label{eqfl}
    \end{gather}
    
    In the proof of Theorem \ref{th1}, recall that for any $ m \in \left\{1,..., N\right\}$, $\phi\left(\theta\right)$ can be divided into two parts which are respectively related to $a_m$ and unrelated to $a_m$. Thus
    \begin{equation}\label{eqphif}
        \phi\left(\theta\right)=f_k \left(\theta_k,\theta_{-k}\right)+\Gamma_{-k},
    \end{equation}
    \begin{equation}
        \phi\left(\theta\right)=f_l \left(\theta_l,\theta_{-l}\right)+\Gamma_{-l}.
    \end{equation}
    Then, $f_l$ must belong to $\Gamma_{-k}$ because $a_k$ and $a_l$ are not neighbors of each other and the terms in $\phi$ belonging to $f_l$ must not belong to $f_k$. In other words, the terms both related to $f_k$ and $f_l$ are all equal to zero. Therefore, $\phi\left(\theta\right)$ can be divided into three parts, which are respectively related to $a_k$, related to $a_l$, and unrelated to both $a_k$ and $a_l$. That is to say,
    \begin{equation*}
        \phi\left(\theta\right)=f_k \left(\theta_k,\theta_{-k}\right)+f_l \left(\theta_l,\theta_{-l}\right)+\Gamma_{-k,-l},
    \end{equation*}
    where $\Gamma_{-k,-l}$ represents the part unrelated to both $a_k$ and $a_l$.
    Consequently,
    \begin{equation*}
        \begin{aligned}
        \phi\left(\theta^{p+1}\right)-\phi\left(\theta^p\right)=&f_k \left(\theta_k^{p+1},\theta_{-k}^{p+1}\right)+f_l \left(\theta_l^{p+1},\theta_{-l}^{p+1}\right)\\
        &-f_k \left(\theta_k^{p},\theta_{-k}^p\right)-f_l \left(\theta_l^{p},\theta_{-l}^p\right)\\
        =&f_k \left(\theta_k^{p+1},\theta_{-k}^{p}\right)-f_k \left(\theta_k^{p},\theta_{-k}^p\right)\\
        &+f_l \left(\theta_l^{p+1},\theta_{-l}^{p}\right)-f_l \left(\theta_l^{p},\theta_{-l}^p\right)\\
        =&\sum_{k\in\mathcal{I}^{p+1}} R_k\left(\theta_k^{p+1},\theta_k^{p},\theta_{-k}^p\right)\\
        =& \sum_{k\in\mathcal{I}^{p+1}} R_k^{p+1}
        \end{aligned}
    \end{equation*}
    where the second equal sign is because of \eqref{eqfk} and \eqref{eqfl}.      
    The above proof can be easily generalized to the case with multiple $\epsilon$-innovators. 
    \hfill $\square$

\section{Proof of Theorem \ref{th2}.}\label{appen2}
Firstly, for any $ \epsilon>0$, we prove by contradiction that there exists $ p \in \left\{ 1,2,...,P\right\}$,  such that $\phi(\theta^p)-\phi\left(\theta^{p-1}\right)<\epsilon$.   Suppose for any $ p\in \left\{1,...,P\right\}$, $\phi(\theta^p)-\phi\left(\theta^{p-1}\right)\geq\epsilon$. On the one hand, one can obtain that 
 \begin{align*}
     \phi(\theta^P)-\phi\left(\theta^{P-1}\right)\geq&~\epsilon,\\
     \phi(\theta^{P-1})-\phi\left(\theta^{P-2}\right)\geq&~\epsilon,\\
     ...\qquad &\\
     \phi(\theta^{1})-\phi\left(\theta^{0}\right)\geq&~\epsilon.
 \end{align*}
Sum the above inequalities, which leads to
$$\phi(\theta^P)-\phi\left(\theta^{0}\right)\geq P\epsilon.$$
On the other hand, according to \eqref{eqP},
\begin{align*}
    \phi(\theta^P)-\phi\left(\theta^{0}\right)\leq& ~ \phi_{max}-\phi_{min}
    <P\epsilon,
\end{align*}
which leads to contradictions. As a result, $\exists~ p_1 \in \left\{1,...,P\right\}$, s.t.,  $\phi(\theta^{p_1})-\phi\left(\theta^{p_1 -1}\right)<\epsilon$.

Secondly, we prove that Algorithm~\ref{DOCS} achieves the $\epsilon$-equilibrium in iteration  $p_1$. Denote $\mathcal{I}^{p_1}$ as the set of indices for all $\epsilon$-innovators in the $p_1$-th iteration. From Lemma \ref{lemma1}, it can be derived that
\begin{equation*}
    \begin{aligned}
        \phi(\theta^{p_1})-\phi\left(\theta^{{p_1}-1}\right)=&\sum_{k\in\mathcal{I}^{p_1}} R_k^{p_1}< \epsilon.
    \end{aligned}
\end{equation*}
Note that according to lines 4 to 9 in Algorithm~\ref{DOCS}, $R_k^{p_1}\geq 0$ holds for all agents.
 It indicates that 
$$\max_k R_k^{p_1} <\epsilon,$$
and, consequently, agents can not gain benefit more than $\epsilon$ by unilaterally improving their strategies. In other words, the $\epsilon$-equilibrium has been reached in the ${p_1}$-th iteration. 

Finally, $\zeta_k$ will be set to 0 for all $k\in\{1,...,N\}$ and all $p\in \{p_1,p_1+1,...,P\}$ according lines 17 to 21 in Algorithm~\ref{DOCS}, which means all agents will not update strategies after the ${p_1}$-th iteration. As a result, Algorithm~\ref{DOCS} will output an $\epsilon$-equilibrium after $P$ iterations.

\hfill $\square$
\end{appendices}





\bibliographystyle{IEEEtran}
\bibliography{reference}

@article{monderer1996potential,
  title={Potential games},
  author={Monderer, Dov and Shapley, Lloyd S},
  journal={Games and economic behavior},
  volume={14},
  number={1},
  pages={124--143},
  year={1996},
  publisher={Elsevier}
}

@article{sun2021game,
  title={Game theoretic self-organization in multi-satellite distributed task allocation},
  author={Sun, Changhao and Wang, Xiaochu and Qiu, Huaxin and Zhou, Qingrui},
  journal={Aerospace Science and Technology},
  volume={112},
  pages={106650},
  year={2021},
  publisher={Elsevier}
}

@inproceedings{karapetyan2017efficient,
  title={Efficient multi-robot coverage of a known environment},
  author={Karapetyan, Nare and Benson, Kelly and McKinney, Chris and Taslakian, Perouz and Rekleitis, Ioannis},
  booktitle={2017 IEEE/RSJ International Conference on Intelligent Robots and Systems (IROS)},
  pages={1846--1852},
  year={2017},
  organization={IEEE}
}

@article{zhang2018fast,
  title={Fast deployment of UAV networks for optimal wireless coverage},
  author={Zhang, Xiao and Duan, Lingjie},
  journal={IEEE Transactions on Mobile Computing},
  volume={18},
  number={3},
  pages={588--601},
  year={2018},
  publisher={IEEE}
}

@article{chen2021general,
  title={A general formal method for manifold coverage analysis of satellite constellations},
  author={Chen, Xiaoyu and Song, Zhiming and Dai, Guangming and Wang, Maocai and Ortore, Emiliano and Circi, Christian},
  journal={IEEE Transactions on Aerospace and Electronic Systems},
  volume={58},
  number={2},
  pages={1462--1479},
  year={2021},
  publisher={IEEE}
}

@article{ulybyshev2008satellite,
  title={Satellite constellation design for complex coverage},
  author={Ulybyshev, Yuri},
  journal={Journal of Spacecraft and Rockets},
  volume={45},
  number={4},
  pages={843--849},
  year={2008}
}

@inproceedings{mannadiar2010optimal,
  title={Optimal coverage of a known arbitrary environment},
  author={Mannadiar, Raphael and Rekleitis, Ioannis},
  booktitle={2010 IEEE International conference on robotics and automation},
  pages={5525--5530},
  year={2010},
  organization={IEEE}
}

@article{sun2023optimal,
  title={Optimal coverage control of stationary and moving agents under effective coverage constraints},
  author={Sun, Xinmiao and Ren, Mingli and Ding, Dawei and Cassandras, Christos G},
  journal={Automatica},
  volume={157},
  pages={111236},
  year={2023},
  publisher={Elsevier}
}

@article{nemer2020game,
  title={A game theoretic approach of deployment a multiple UAVs for optimal coverage},
  author={Nemer, Ibrahim A and Sheltami, Tarek R and Mahmoud, Ashraf S},
  journal={Transportation Research Part A: Policy and Practice},
  volume={140},
  pages={215--230},
  year={2020},
  publisher={Elsevier}
}

@article{xiao2020distributed,
  title={A distributed multi-agent dynamic area coverage algorithm based on reinforcement learning},
  author={Xiao, Jian and Wang, Gang and Zhang, Ying and Cheng, Lei},
  journal={IEEE Access},
  volume={8},
  pages={33511--33521},
  year={2020},
  publisher={IEEE}
}

@article{gao2022coverage,
  title={Coverage control for UAV swarm communication networks: A distributed learning approach},
  author={Gao, Ning and Liang, Le and Cai, Donghong and Li, Xiao and Jin, Shi},
  journal={IEEE Internet of Things Journal},
  volume={9},
  number={20},
  pages={19854--19867},
  year={2022},
  publisher={IEEE}
}

@article{wang2024distributed,
  title={Distributed learning-based visual coverage control of multiple Mobile Aerial Agents},
  author={Wang, Ye and Fu, Junjie and Tang, Meiqi},
  journal={Journal of the Franklin Institute},
  volume={361},
  number={5},
  pages={106683},
  year={2024},
  publisher={Elsevier}
}

@article{meng2021deep,
  title={Deep reinforcement learning-based effective coverage control with connectivity constraints},
  author={Meng, Shaofeng and Kan, Zhen},
  journal={IEEE Control Systems Letters},
  volume={6},
  pages={283--288},
  year={2021},
  publisher={IEEE}
}

@article{yazicioglu2016communication,
  title={Communication-free distributed coverage for networked systems},
  author={Yaz{\i}c{\i}oglu, A Yasin and Egerstedt, Magnus and Shamma, Jeff S},
  journal={IEEE Transactions on Control of Network Systems},
  volume={4},
  number={3},
  pages={499--510},
  year={2016},
  publisher={IEEE}
}

@article{dong2023energy,
  title={Energy-Efficient Sensor Deployment Strategy for Optimal Coverage of Underwater Events Inspired by Krill Herd},
  author={Dong, Mingru and Li, Haibin and Li, Cheng and Hu, Yongtao and Huang, Haocai},
  journal={IEEE Transactions on Industrial Informatics},
  year={2023},
  publisher={IEEE}
}

@article{huang2022deployment,
  title={Deployment of heterogeneous UAV base stations for optimal quality of coverage},
  author={Huang, Hailong and Savkin, Andrey V},
  journal={IEEE Internet of Things Journal},
  volume={9},
  number={17},
  pages={16429--16437},
  year={2022},
  publisher={IEEE}
}

@article{hu2023multi,
  title={Multi-UAV coverage path planning: A distributed online cooperation method},
  author={Hu, Wenjian and Yu, Yao and Liu, Shumei and She, Changyang and Guo, Lei and Vucetic, Branka and Li, Yonghui},
  journal={IEEE Transactions on Vehicular Technology},
  volume={72},
  number={9},
  pages={11727--11740},
  year={2023},
  publisher={IEEE}
}

@article{varposhti2020distributed,
  title={Distributed coverage in mobile sensor networks without location information},
  author={Varposhti, Marzieh and Hakami, Vesal and Dehghan, Mehdi},
  journal={Autonomous Robots},
  volume={44},
  number={3},
  pages={627--645},
  year={2020},
  publisher={Springer}
}

@article{trotta2018joint,
  title={Joint coverage, connectivity, and charging strategies for distributed UAV networks},
  author={Trotta, Angelo and Di Felice, Marco and Montori, Federico and Chowdhury, Kaushik R and Bononi, Luciano},
  journal={IEEE Transactions on Robotics},
  volume={34},
  number={4},
  pages={883--900},
  year={2018},
  publisher={IEEE}
}

@article{sun2017time,
  title={A time variant log-linear learning approach to the SET K-COVER problem in wireless sensor networks},
  author={Sun, Changhao},
  journal={IEEE transactions on cybernetics},
  volume={48},
  number={4},
  pages={1316--1325},
  year={2017},
  publisher={IEEE}
}

@article{abdulghafoor2021two,
  title={Two-level control of multiagent networks for dynamic coverage problems},
  author={Abdulghafoor, Alaa Z and Bakolas, Efstathios},
  journal={IEEE Transactions on Cybernetics},
  volume={53},
  number={7},
  pages={4067--4078},
  year={2021},
  publisher={IEEE}
}

@article{chen2023vleo,
  title={VLEO satellite constellation design for regional aviation and marine coverage},
  author={Chen, Guoquan and Wu, Shaohua and Deng, Yajing and Jiao, Jian and Zhang, Qinyu},
  journal={IEEE Transactions on Network Science and Engineering},
  volume={11},
  number={1},
  pages={1188--1201},
  year={2023},
  publisher={IEEE}
}

@article{wu2024joint,
  title={Joint power and coverage control of massive UAVs in post-disaster emergency networks: An aggregative game-theoretic learning approach},
  author={Wu, Jing and Chen, Qimei and Jiang, Hao and Wang, Haozhao and Xie, Yulai and Xu, Wenzheng and Zhou, Pan and Xu, Zichuan and Chen, Lixing and Li, Beibei and others},
  journal={IEEE Transactions on Network Science and Engineering},
  volume={11},
  number={4},
  pages={3782--3799},
  year={2024},
  publisher={IEEE}
}

@article{ai2008optimality,
  title={Optimality and complexity of pure Nash equilibria in the coverage game},
  author={Ai, Xin and Srinivasan, Vikram and Tham, Chen-khong},
  journal={IEEE Journal on Selected Areas in Communications},
  volume={26},
  number={7},
  pages={1170--1182},
  year={2008},
  publisher={IEEE}
}

@article{li2017potential,
  title={A potential game approach to multiple UAV cooperative search and surveillance},
  author={Li, Pei and Duan, Haibin},
  journal={Aerospace Science and Technology},
  volume={68},
  pages={403--415},
  year={2017},
  publisher={Elsevier}
}

@article{zhang2024distributed,
  title={Distributed dynamic task allocation for unmanned aerial vehicle swarm systems: A networked evolutionary game-theoretic approach},
  author={Zhang, Zhe and Jiang, Ju and ZHANG, Wen-An and others},
  journal={Chinese Journal of Aeronautics},
  volume={37},
  number={6},
  pages={182--204},
  year={2024},
  publisher={Elsevier}
}

@article{yang2013towards,
  title={Towards a snowdrift game optimization to vertex cover of networks},
  author={Yang, Yang and Li, Xiang},
  journal={IEEE Transactions on Cybernetics},
  volume={43},
  number={3},
  pages={948--956},
  year={2013},
  publisher={IEEE}
}

@ARTICLE{10876799,
  author={Jiang, Bin and Zhou, Ronghao and Luo, Fei and Cui, Xuerong and Wang, Huihui Helen and Song, Houbing Herbert},
  journal={IEEE Transactions on Network Science and Engineering}, 
  title={Attack Detection and Optimal Deployment for Underwater Constrained Wireless Sensor Networks via Hybrid Trust Evidence}, 
  year={2025},
  volume={12},
  number={3},
  pages={1791-1803},
}

@article{ma2023dynamic,
  title={Dynamic tracking coverage with quantity-adjustment behavior},
  author={Ma, Shizhuo and Li, Shengjin and Yu, Dengxiu and Wang, Zhen and Liu, Yan-Jun and Chen, CL Philip},
  journal={IEEE Transactions on Systems, Man, and Cybernetics: Systems},
  volume={53},
  number={10},
  pages={6094--6106},
  year={2023},
  publisher={IEEE}
}

@article{alzenad20173,
  title={3-D placement of an unmanned aerial vehicle base station (UAV-BS) for energy-efficient maximal coverage},
  author={Alzenad, Mohamed and El-Keyi, Amr and Lagum, Faraj and Yanikomeroglu, Halim},
  journal={IEEE Wireless Communications Letters},
  volume={6},
  number={4},
  pages={434--437},
  year={2017},
  publisher={IEEE}
}

@article{zhao2021multi,
  title={Multi-UAV trajectory planning for energy-efficient content coverage: A decentralized learning-based approach},
  author={Zhao, Chenxi and Liu, Junyu and Sheng, Min and Teng, Wei and Zheng, Yang and Li, Jiandong},
  journal={IEEE Journal on Selected Areas in Communications},
  volume={39},
  number={10},
  pages={3193--3207},
  year={2021},
  publisher={IEEE}
}

@ARTICLE{9904913,
  author={Yengejeh, Armin Sadeghi and Asghar, Ahmad Bilal and Smith, Stephen L.},
  journal={IEEE Transactions on Control of Network Systems}, 
  title={Distributed Multirobot Coverage Control of Nonconvex Environments With Guarantees}, 
  year={2023},
  volume={10},
  number={2},
  pages={796-808},
  doi={10.1109/TCNS.2022.3210328}}

@ARTICLE{10829684,
  author={Zou, Jiayu and Zhang, Hai-Tao and Zhai, Chao and Xing, Ning and Ma, Yong and Liu, Xingjian},
  journal={IEEE Transactions on Control of Network Systems}, 
  title={Cooperative Hierarchical Coverage Control of Multiagent Systems With Nonholonomic Constraints in Poriferous Environments}, 
  year={2025},
  volume={12},
  number={2},
  pages={1512-1520},
  doi={10.1109/TCNS.2025.3526566}}
\end{document}